\documentclass[floatfix,preprint,aps,prd,showpacs,amsmath,amssymb]{revtex4}
\usepackage{graphicx}
\usepackage{epsfig}
\usepackage{psfrag}
\newcommand{\be}{\begin{equation}}
\newcommand{\ee}{\end{equation}}

\newcommand{\bh}{\{B_k\}}
\newcommand{\ba}{\mbox{\boldmath$a$}}
\newcommand{\df}{\Delta f}
\newcommand{\e}{{\rm e}}
\newcommand{\rd}{\,{\rm d}}
\renewcommand{\vec}[1]{\mathbf{#1}}
\begin{document}


\title{A Metropolis-Hastings algorithm for extracting periodic
gravitational wave signals from laser interferometric detector data}


\author{Nelson Christensen$^1$\footnote{nchriste@carleton.edu},
R\'ejean J. Dupuis$^2$\footnote{rejean@astro.gla.ac.uk},
Graham Woan$^2$\footnote{graham@astro.gla.ac.uk}
and Renate Meyer$^3$\footnote{meyer@stat.auckland.ac.nz}}
\affiliation{$^1$Physics and Astronomy, Carleton College,
Northfield, MN 55057, USA\\
$^2$Department of Physics and Astronomy, University of Glasgow, G12 8QQ, United Kingdom \\
$^3$Department of Statistics, University of Auckland,
Auckland, New Zealand}

\date{\today}

\begin{abstract}
The Markov chain Monte Carlo methods offer practical procedures
for detecting signals characterized by a large number of
parameters and under conditions of low signal-to-noise ratio. We
present a Metropolis-Hastings algorithm capable of inferring the
spin and orientation parameters of a neutron star from its
periodic gravitational wave signature seen by laser
interferometric detectors.
\end{abstract}


\maketitle

\section{Introduction}
The world-wide network of laser interferometric gravitational wave
detectors has begun to acquire scientifically significant data
\cite{ligo,geo,virgo,tama} and rapidly rotating neutron stars are
an important potential source of signals (we will reserve the term
`pulsar' to refer to the observed pulsating radio sources).  Although a
spinning spherically symmetric neutron star will not produce
gravitational waves, a number of mechanisms have been proposed
that are capable of producing quasi-periodic gravitational waves
from biaxial or triaxial neutron stars \cite{Cutler02,bildsten}. 
Any gravitational waves from these neutron stars will
likely be seen at Earth as weak continuous wave signals.

The data analysis task of identifying such a signal in the output
of a laser interferometer is challenging and difficult, both
because of the weakness of the signal and because its time
evolution is characterised by a relatively large number of
parameters.  Radio observations can provide the sky location,
rotation frequency and spindown rate of known pulsars, but the
problem of looking for unknown (or poorly parameterised) neutron
star sources is significantly more challenging. SN1987A is a good
example of a poorly parameterised source for which the sky
location in approximately known but also for which there is a
large uncertainty in the frequency and spindown parameters of the
putative neutron star \cite{midd}.

Much work has already gone into all-sky hierarchical methods for
searching for continuous gravitational waves \cite{jaranowski,stack}.  Here we
address the specific problem of a `fuzzy' parameter space search,
in which a restricted volume of the space needs to be thoroughly
investigated. We take a Bayesian approach to this problem and use
Markov chain Monte Carlo (MCMC) techniques which have been  shown
to be especially suited to similar problems involving numerous
parameters \cite{gilks96}. In particular, the Metropolis-Hastings
(MH) algorithm \cite{metr53,hastings} has been used for estimating
cosmological parameters from cosmic microwave background data
\cite{nlc1,knox,spergel}, and the applicability of the MH routine
has been demonstrated in estimating astrophysical parameters for
gravitational wave signals from coalescing compact binary systems
\cite{nlc2,nlc3}. MCMC methods have also provided Bayesian
inference for noisy and chaotic data \cite{meyer1,meyer2}.

Here we demonstrate that a MH algorithm also offers great promise
for estimating neutron star parameters from their continuous
gravitational wave signals. This works builds on the development
(by two of us) of an end-to-end robust Bayesian method of
searching for periodic signals in gravitational wave
interferometer data \cite{Dupuis}, summarized in
Sec.~\ref{signal}. Starting with this Bayesian approach we apply a
similar MH routine to that used in \cite{nlc1,nlc3}. The
description of the Bayesian MH method is described in
Sec.~\ref{TMHA}. In Sec.~\ref{results} we present the results of
this study, using synthesized data, for four and five parameter
problems. We believe that this method offers great hope for signal
extraction as more parameters are included, and this point is
discussed in Sec.~\ref{disc}.

\section{Signal Characteristics}
\label{signal}
 We will initially consider searching for signals from
known radio pulsars, and then expand the method to account for an
uncertainty in the frequency of the gravitational wave signal. As
gravitational waves from pulsars are certainly weak at Earth, long
integration periods are required to extract the signal, and we
must take account of the antenna patterns of the detectors and the
Doppler shift due to the motion of the Earth.

As in the previous study \cite{Dupuis,lscCW} we consider the
signal expected from a non-precessing triaxial neutron star. The
gravitational wave signal from such an object is at twice its
rotation frequency, $f_{\rm s}=2 f_{\rm r}$, and we characterise
the amplitudes of each polarization with overall strain factor,
$h_0$. The measured gravitational wave signal will also depend on
the polarisation antenna patterns of the detector $F_{\times, +}$
giving a signal
\begin{equation}
s(t) = \frac{1}{2}F_{+}(t;\psi)h_{0}(1 + \cos^{2}\iota)\cos
\Psi(t) + F_{\times} (t;\psi)h_{0}\cos \iota \sin  \Psi(t),
\label{s}
\end{equation}
where $\psi$ is the polarization angle of the radiation (which
depends on the position angle of the spin axis in the plane of the
sky) and $\iota$ is the inclination of the pulsar with respect to
the line-of-sight.

Using a simple slowdown model, the phase evolution of the signal
can be usefully parameterised as
\begin{equation}
 \Psi(t) = \phi_{0} + 2\pi \left[f_{\rm s}(T - T_{0}) + \frac{1}{2}\dot{f_{\rm s}} (T -
       T_{0})^{2} + \frac{1}{6}\ddot{f_{\rm s}}(T - T_{0})^{3}\right],
\label{phase1}
\end{equation}
where
\begin{equation}
T = t + \delta t= t + \frac{\vec{r} \cdot \vec{n}}{c}  +
\Delta_{T}. \label{time}
\end{equation}
Here, $T$ is the time of arrival of the signal at the solar system
barycenter, $\phi_{0}$ is the phase of the signal at a fiducial
time $T_{0}$, $\vec{r}$ is the position of the detector with
regard to the solar system barycenter, $\vec{n}$ is a unit vector
in the direction of the pulsar, $c$ is the speed of light, and
$\Delta_{T}$ contains the relativistic corrections to the arrival
time \cite{taylor}.

The signal is {\it heterodyned} by multiplying the data by $\exp(-i\Psi(t))$ 
so that the only time varying quantity remaining is the antenna pattern of the 
interferometer (which varies over the day). For convenience, the 
result is low-pass filtered and resampled.
We are left with a
simple model with four unknown parameters: the overall amplitude
of the gravitational wave signal ($h_0$), its polarization angle
($\psi$), its phase at time $T_{0}$ ($\phi_{0}$) , and the angle
between the spin axis of the pulsar and the line-of-sight
($\iota$).

A detailed description of the heterodyning procedure is presented
elsewhere \cite{Dupuis, lscCW}; here we just provide a summary of
this standard technique. The raw signal, $s(t)$, is centered near
twice the rotation frequency of the pulsar, but is Doppler
modulated due to the motion of the Earth and the orbit of the
pulsar if it is in a binary system. The modulation bandwidth is
typically $~10^4$ times less than the detector bandwidth, so one
can greatly reduce the effective data rate by extracting this band
and shifting it to zero frequency.  In its standard form the
result is one binned data point, $B_{k}$, every minute, containing
all the relevant information from the original time series but at
only $2\times10^{-6}$ the original data rate.  If the phase evolution has
been correctly accounted for at this heterodyning stage then the
only time-varying component left in the signal will be the effect
of the antenna pattern of the interferometer, as its geometry with
respect to the neutron star varies with Earth rotation.   Any
small error, $\df$, in the heterodyne frequency will cause the
signal to oscillate at  $\df$, and for the second part of our
study we have $\df$ as our fifth parameter. For both these studies
we estimate the noise variance, $\sigma^2_k$, in the bin values,
$B_{k}$, from the sample variance of the contributing data. It is
assumed that the noise is stationary over the 60\,s of data
contributing to each bin.

\section{The Metropolis-Hastings Algorithm}
\label{TMHA}
 This section presents a brief review of the Bayesian MH
approach to parameter estimation. Comprehensive descriptions of
MCMC methods and the MH algorithm can be found elsewhere
\cite{gilks96,nlc1,nlc3}.

We will denote the output from the above heterodyning procedure as
$\bh$, with joint probability distribution function (pdf)
denoted by $p(\bh|\ba)$ conditional on unobserved parameters
$\ba=(a _1,\ldots , a _d)$. The pdf $p(\bh|\ba)$ is referred to as
the \emph{likelihood} and regarded as a function of the parameters
$\ba$. The parameters of interest for our four parameter study are
$\ba=(h_0,\psi,\phi_0,\iota)$, while for the five parameter study
they are $\ba=(h_0,\psi,\phi_0,\iota,\df)$.

From Eq.~(\ref{s}), the (now complex) heterodyned signal is
\begin{equation}
 y(t_k;\ba) = \frac{1}{4}F_{+}(t_k;\psi)h_{0} (1 +
 \cos^{2}\iota)\e^{i\phi_{0}}
 - \frac{i}{2}F_{\times}(t_k;\psi) h_{0} \cos\iota
 \e^{i\phi_{0}},
\end{equation}
and the binning procedure should, by the central limit theorem,
give the noise a near-gaussian probability density characterized
by a variance $\sigma^2_k$ for the $k$th bin. The likelihood that
the data in this bin, taken at time $t_k$, is consistent with the
above model is
 \begin{equation}
 p(B_k|\ba)\propto\exp\left(\frac{-|B_k-y(t_k;{\ba})|^2}{2\sigma_k^2}\right),
\end{equation}
and the joint likelihood that the data in all the bins (taken as
independent) are consistent with a particular set of model
parameters is
\begin{eqnarray}
 p(\bh|\ba)\propto\prod_k \exp\left(\frac{-|B_k-y(t_k;\ba)|^2}{2\sigma_k^2}\right).
\end{eqnarray}

Bayesian inference requires the specification of a prior pdf for
$\ba$, $p(\ba)$, that quantifies the researcher's pre-experimental
knowledge about $\ba$. The phase and polarisation priors are flat
in their space, and are set uniform for $\phi_{0}$ over $[0,\pi]$,
and for $\psi$ over $[-\pi/4,\pi/4]$. The prior for $\iota$ is
uniform in $\cos\iota$ over $[-1,1]$, corresponding to a uniform
prior per unit solid angle of pulsar orientation. Finally, in the
present study we take a prior for $h_0$ that is uniform for
$0<h_0<1000$ (in our normalized units for which $\sigma_k=1$), and
zero for all other values.

Using Bayes' theorem, the post-experimental knowledge of $\ba$ is
expressed by the \emph{posterior} pdf of $\ba$:
 \begin{equation}
p(\ba|\bh)= \frac{p(\ba)p(\bh|\ba)}{p(\bh)}\propto
p(\ba)p(\bh|\ba),
 \end{equation}
where $p(\bh)=\int p(\bh|\ba)p(\ba)\rd\ba$ is the marginal pdf of
$\bh$ which can be regarded as a normalizing constant as it is
independent of $\ba$. The posterior pdf is thus proportional to
the product of prior and likelihood.

The marginal posterior distribution for parameter $a_i$ is the
integral of the joint posterior pdf over all other components of
$\ba$ other than $ a_{i}$, i.e.,
\begin{equation}
p(a_{i}|\bh)=\int\ldots\int p(\ba|\bh )\rd a_{1}\ldots \rd
a_{i-1}\rd a_{i+1}\ldots \rd a_{d},
\end{equation}
and contains all the analysis has to say about the value of $a_i$
alone.  However it is often useful to summarise this in a single
`point estimate' of $a_i$ using, for example, the posterior mean:
\begin{equation}\label{eq:mean}
\langle a_{i}\rangle = \int a_{i}p( a_{i}|\bh)\rd a_{i}.
\end{equation}
Calculating the normalization constant $p(\bh)$ and calculating
each marginal posterior pdf requires difficult $d$- and $d-1$
dimensional  integrations, respectively, that can be evaluated using a sampling
approach and MCMC methods \cite{gilks96,nlc1,nlc3}. Rather than
sampling directly from $p(\ba|\bh)$, a sample from a Markov chain
is generated which has $p(\ba|\bh)$ as its equilibrium
distribution. Thus, after running the Markov chain for a certain
`burn-in' period, these (correlated) samples can be regarded as
samples from the limiting distribution, provided that the Markov
chain has reached convergence. Despite their correlations, the
ergodic theorem guarantees that the sample average is still a
consistent estimate of the posterior mean Eq.~(\ref{eq:mean}) \cite{tierney}.

The specific MCMC technique used for this study was the MH
algorithm \cite{metr53,hastings}. The MH algorithm generates a
sample from the target pdf $p( \ba|\bh )$ using a technique that
is similar to the well-known simulation technique of
\emph{rejection sampling}. A candidate is generated from an
auxiliary pdf and then accepted or rejected with some probability.
Starting with an arbitrary initial state $\ba_0$, at time $n$ a
new candidate $\ba'$ is generated from the candidate generating
pdf, $q(\ba|\ba_{n})$, which can depend on the current state
$\ba_{n}$ of the Markov chain. This new candidate $ \ba'$ is
accepted with a certain \emph{acceptance probability} $\alpha
(\ba'| \ba_{n})$, also depending on the current state $\ba_{n}$,
given by
\be
\alpha(\ba'|\ba_n)=\min \left\{\frac{p(\ba')p(\bh|\ba')q(
\ba_n|\ba')}{p(\ba_n)p(\bh|\ba_n)q(\ba'|\ba_n)},1\right\}.
\ee
For good efficiency a multivariate normal distribution
centered at the current state $\ba_n$
is used for
$q(\ba'|\ba_n)$. This then implies that if the posterior
probability at  $\ba'$ is larger than at the current state
$\ba_{n}$, the proposed step to $\ba'$ is always accepted.
However, if the step is in a direction of lower posterior
probability, then this step is accepted only with a certain
probability given by the ratio of the posterior pdfs (since our
multivariate normal generating function is symmetric in $\ba'$ and
$\ba_n$ and therefore cancels out). If the candidate is accepted,
the next state of the Markov chain is $\ba_{n+1}= \ba'$, otherwise
the chain does not move, i.e. $\ba_{n+1}= \ba_{n}$.

The steps of the MH algorithm are therefore:
\begin{center}
\begin{tabular}{rl}
 Step 0:     & Start with an arbitrary value $\ba_0$ \\[2pt]
 Step $n+1$: & Generate $\ba'$ from $q(\ba|\ba_{n})$ and $u$ from $U(0,1)$ \\
             & If $u\leq\alpha(\ba'| \ba_n)$ set $\ba_{n+1}=\ba'$ (acceptance) \\
             & If $u>\alpha(\ba'|\ba_n)$ set $\ba_{n+1}=\ba_n$ (rejection).
\end{tabular}
\end{center}
The efficiency of the MH algorithm depends heavily on the choice
of the proposal density. The closer the proposal is to the target
distribution, the faster convergence will be accomplished. This
link between the closeness of the proposal to stationary
distribution and speed of convergence has also been substantiated
by Holden \cite{hold98}. In the study presented here we
dynamically altered the proposal distribution based on information
from the chain's history. The approach, called \emph{pilot
adaptation}, is to perform a separate pilot run to gain insight
about the target density and then tune the proposal accordingly
for the successive runs. Such adaptation can be iterated but
allowing it infinitely often will destroy the Markovian property
of the chain and thereby often compromise the stationarity of the
chain and the consistency of sample path averages (\cite{gilk98};
see \cite{gelf94} for an example).

Based on the central limit theorem, the posterior pdf should be
well approximated by a multivariate normal distribution with mean
equal to the posterior mode and covariance matrix equal to minus
the Hessian evaluated at the posterior mode. Thus, we use a
multivariate normal distribution for the proposal density $q(\ba
|\ba_{n})$. As the mode is unknown, we try to make use of pilot
samples to estimate its covariance matrix. When we initially run
the MH algorithm, we sample candidate parameters from a normal
distribution with covariance matrix equal to the identity matrix
and centered around the current state. After the completion of
this pilot run we use the empirical covariance matrix of the
sample as covariance matrix of the multivariate normal proposal
density, again with mean equal to the current state.

\section{Results}
\label{results} In the first part of our study we reproduced the
results presented in \cite{Dupuis} where the four unknown
parameters were $h_0, \iota, \psi$, and $\phi_{0}$. The signal
$s(t)$ was synthesized assuming a source at
$\text{RA}=4^{\rm h} 41^{\rm m} 54^{\rm s}$ and $\text{dec}=18^\circ \, 23' \, 32''$,
as would be seen by the LIGO-Livingston interferometer.
This was then added to white gaussian noise, $n(t)$,
which is a good approximation to the detector noise in our band.
Our normalized data had a noise variance of $\sigma_k^2=1$ for
each sample, and the amplitude of the signal used in our test runs
was varied in the range $h_0=0.0$ to $10.0$. We were able to
detect signals for $h_0 > 0.1$. The length of the data set
corresponded to $14\,400$ samples or 10 days of data at a rate of
one sample per minute (which was the rate used for the LIGO/GEO S1
analysis described in \cite{lscCW}). Although we will work with
strains normalised to $\sigma_k=1$, the results can be cast into a
more conventional form by multiplying $\sigma_i$ and $h_0$ by
$(S_{h}/60)^{1/2}$, where $(S_h)^{1/2}$ is the strain noise
spectral density of the detector at the frequency of interest, in
Hz$^{-1}$.

\begin{figure}
\psfig{file=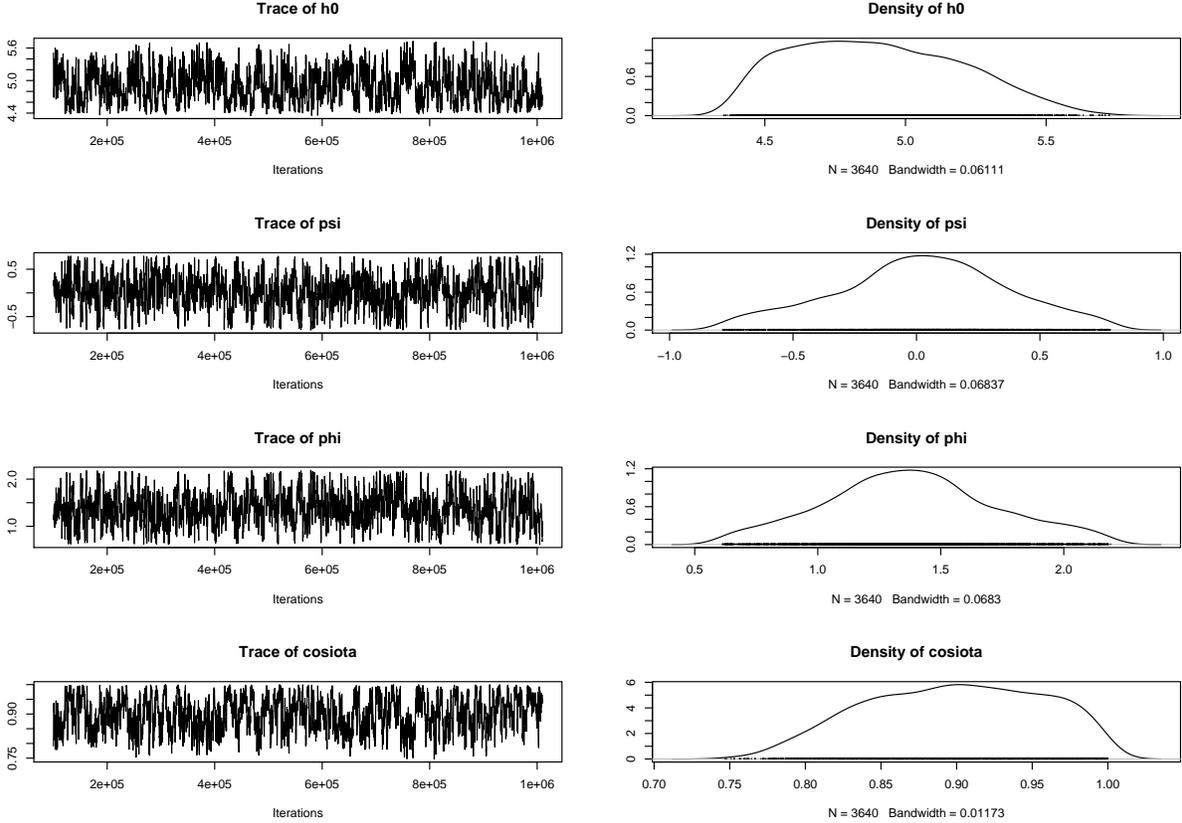,angle=-90,width=1.0 \linewidth,clip=}
\caption{\label{f1} Trace plot (left) and MCMC-estimated posterior
pdfs (right) for the pulsar parameters $h_0$, $\psi$, $\phi_0$ and
$\cos\iota $. In this example the true parameters were $h_0=5.0$,
$\phi_0=1.0$, $\psi=0.4$, and $\iota=0.5$ implying $\cos\iota
=0.88$. }
\end{figure}

\begin{figure}
\includegraphics[ width=5in]{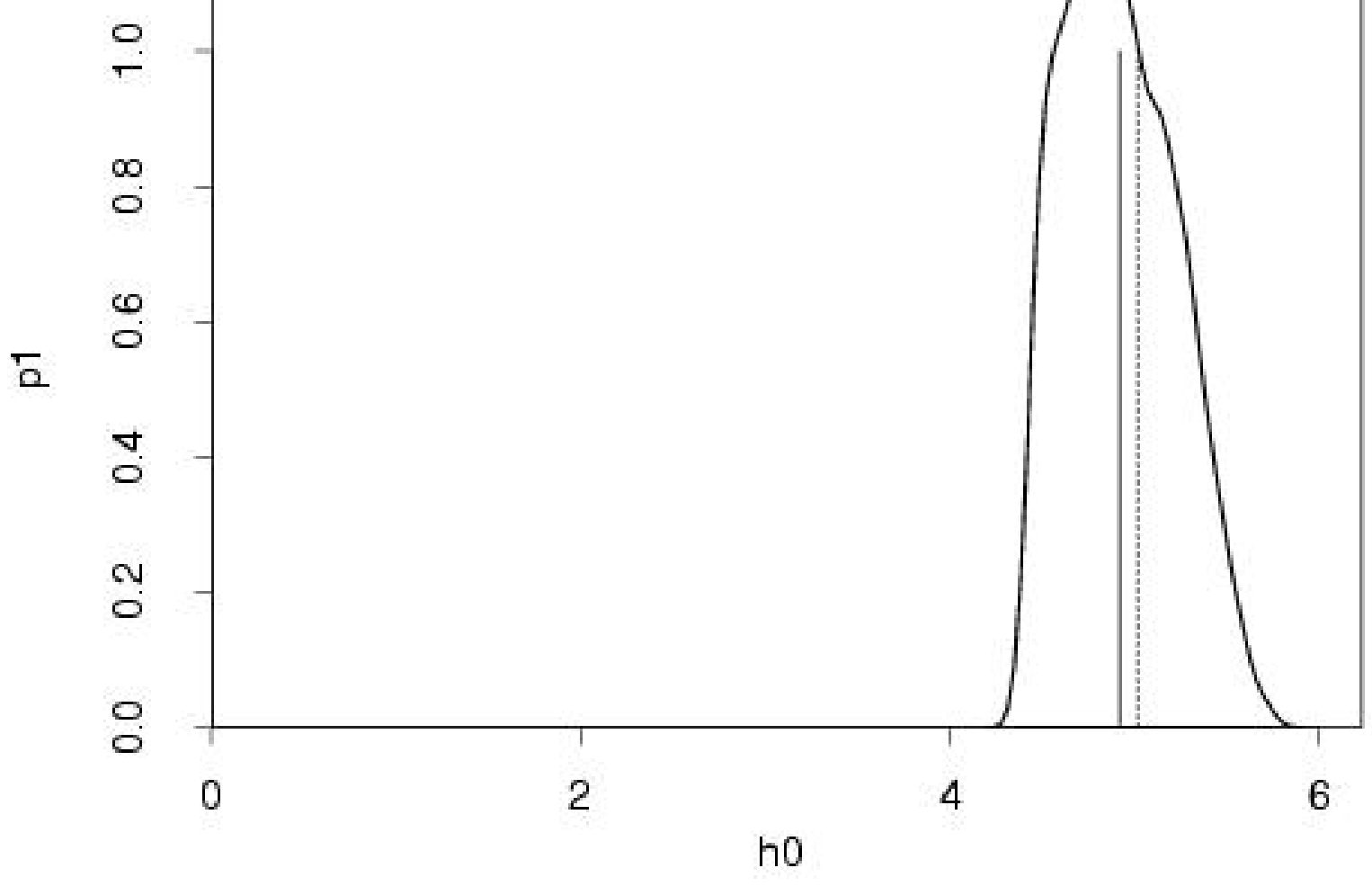}
\caption{\label{fa} An expanded view of the estimated posterior
pdf based on the MCMC sample for parameter $h_0$. The vertical
solid line shows the posterior mean of $h_0=4.9$, while the
vertical dotted line marks the true parameter value of $h_0=5.0$.
}
\end{figure}

An example of the MH routine output is shown in Fig.~\ref{f1}.
Displayed are the trace plots and the kernel densities (posterior
pdfs). For this example the program ran for $10^6$ iterations. The
first $10^5$ iterations were discarded as the burn-in. Short-term
correlations in the chain were eliminated by `thinning' the
remaining terms; we kept every $250^{\rm th}$ item in the chain.
The true parameter values for this run were $h_0=5.0$, $\psi=0.4$,
$\phi_0=1.0$ and $\iota=0.5$ ($\cos\iota =0.88$). In the example
displayed in Fig.~\ref{f1} the MCMC yielded mean values and 95\%
posterior probability intervals of $h_0=4.9$ ($4.43$ to $5.50$),
$\psi=0.02$ ($-0.68$ to $0.69$), $\phi_0=1.34$ ($0.71$ to $2.08$),
and $\cos\iota =0.90$ ($0.79$ to $0.99$). The 95\% posterior
probability interval is specified by the 2.5\% and 97.5\%
percentile of $p( a_{i}|\bh)$. In Fig.~\ref{fa} we display the
estimated posterior pdf of $h_0$ on an expanded scale, along with
the real and estimated value for $h_0$.

\psfrag{p1}{$p(h_0)$} \psfrag{h0}{$h_0$}

It is crucial that our algorithm is sensitive to the true value of
the gravitational wave amplitude, $h_0$, even under conditions of
relatively low signal-to-noise ratio, and Fig.~\ref{f2} shows
injected $h_0$ values versus their values inferred by the MH
routine. The error bars correspond to the 95\% posterior
probability interval, i.e. the lower and upper bound are specified
by the 2.5\% and 97.5\% percentile of $p( a_{i}|\bh)$. The
algorithm clearly is successful in finding and estimating $h_0$.
While the error bars increase as the signal gets larger, the
relative error $\Delta\!h_0 / h_0$ does diminish as $h_0$
increases. The fact that the 95\% posterior probability interval
grows with $h_0$ for constant noise level would seem to be
counterintuitive. In addition, the widths of the posterior
probability distributions for $h_0$ are larger than would be
naively expected from a search for a simple periodic signal.
The reason is that these error bars represent the uncertainty in
the parameter rather than just the level of the noise, and this
is affected both by the noise level and the posterior covariance
between all of the parameters.  The MCMC technique also allows one to
calculate cross-correlation coefficients from the Markov chains of
the parameters, and the value between $h_0$ and $\cos\iota $ in
all of our runs was $\sim-0.95$. As a result the
data are consistent with a relatively broad range of combinations
of the two parameters, making their \emph{individual} values
rather uncertain here -- an effect evident from Eq.~(\ref{s}).

{\psfrag{MCMC}{MCMC-derived $h_0$} \psfrag{Actual}{true $h_0$}
\psfrag{0}{0}\psfrag{1}{1}\psfrag{2}{2}\psfrag{3}{3}\psfrag{4}{4}
\psfrag{6}{6}\psfrag{8}{8}\psfrag{10}{10} \psfrag{0.0}{0.0}
\psfrag{0.5}{0.5} \psfrag{1.0}{1.0} \psfrag{1.5}{1.5}
\begin{figure}
\includegraphics[width=5in]{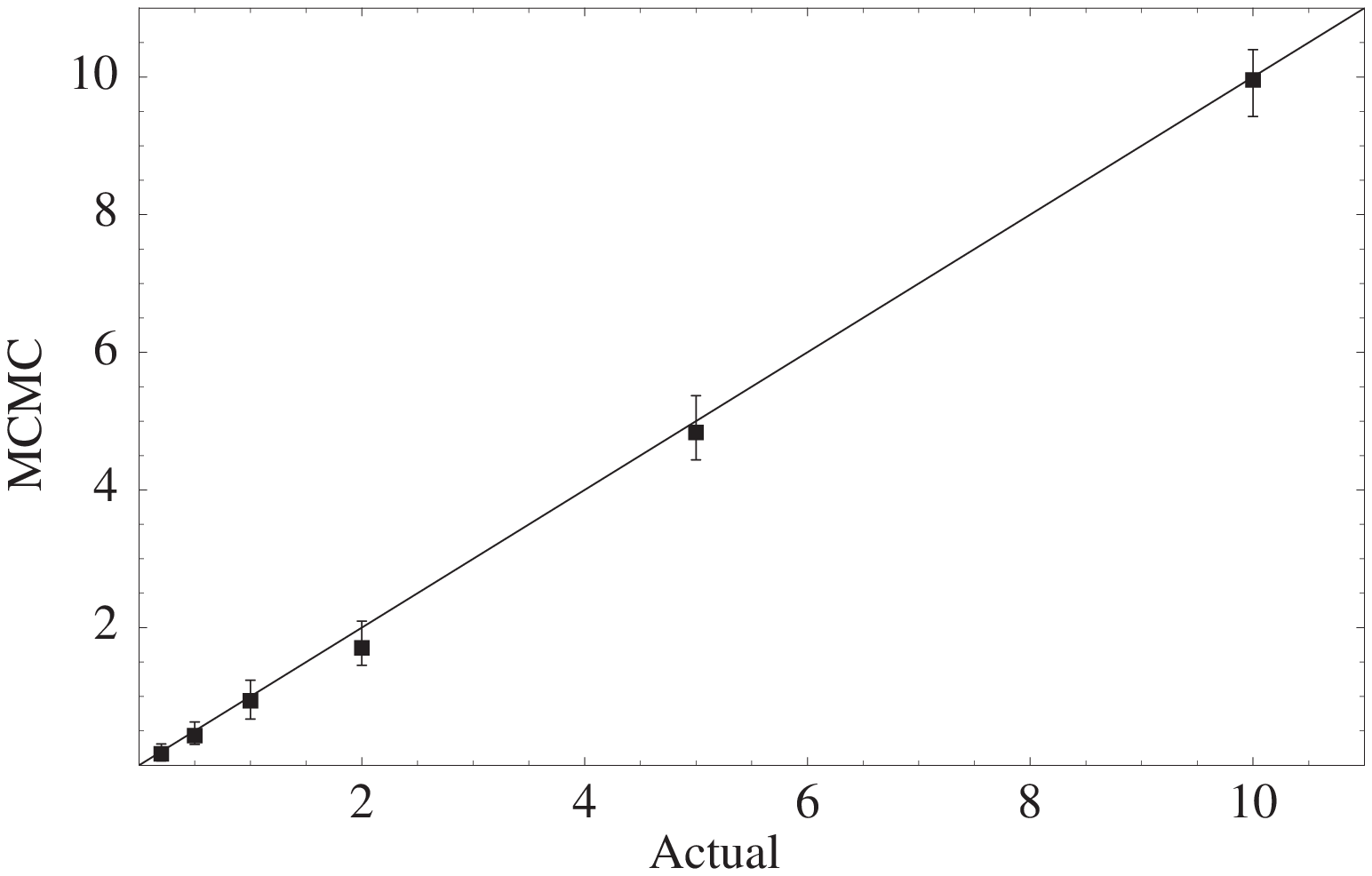}
\caption{\label{f2} The posterior mean based on the MCMC
sample for the gravitational wave amplitude parameter $h_0$ versus
the actual value of $h_0$ used in synthesizing the data. The error
bars correspond to lower and upper bounds at the 2.5\% and 97.5\%
percentiles of the posterior pdf. The solid line has a slope of
$1$. The calculations were performed over 14\,000 data points,
each with noise variance of $\sigma_k^2=1$. }
\end{figure}
}

The effect of the other unknown parameters (particularly $\iota$)
on the posterior pdf for $h_0$ can be clearly shown by repeating
the analysis for Fig.~\ref{f2} but with $h_0$ as the only unknown,
namely, all of the other parameters set to their {\it actual}
values in the MCMC routine. Under these circumstances the widths
of all 95\% posterior probability intervals are $0.116$,
independent of the value of $h_0$. Comprehensive analyses have
investigated detection statistics for a periodic signal in a
gravity wave detector \cite{Allen02}. However, these statistics
are concerned only with the amplitude of the periodic signal, and
not with parameter estimation (as described above). If we write
Eq.~\ref{s} as $s(t)=A \cos(\Psi + \Phi)$ (with $A$ being the
periodic signal amplitude and $\Phi$ a phase term) then the
detection statistic of \cite{Allen02} would apply to finding a
signal amplitude $A$ in the presence of the detector noise. In
terms of Eq.~\ref{s}, the amplitude of the periodic signal would
be
\begin{equation}
A = \{[F_{+}(t;\psi)h_{0}(1 + \cos^{2}\iota)/2]^{2}
+ [F_{\times} (t;\psi)h_{0}\cos \iota]^{2}\}^{1/2}.
\label{amp}
\end{equation}
It is clear that $A$ has a complicated dependence on
$h_0$ and $\cos\iota$. We will never know,
{\it a priori}, the value of all the pulsar parameters. Our
study here is about parameter estimation, and not knowing
the values of all the pulsar parameters ultimately increases the
width in the posterior pdf for the
gravity wave magnitude $h_0$.

As the magnitudes of the signals are diminished there comes a point
when one is no longer able to {\it confidently} claim a detection.
This threshold is somewhat arbitrary, and dependent on the statistics and interpretation.
In the study presented here we {\it claim} that a signal is detected
when the $h_0=0$ point is more that two standard deviations from the mean
value of the MCMC generated posterior pdf for $h_0$. For the
synthesized signals we investigated this corresponded to
a threshold for detection of $h_0=0.1$; in this case the measured
mean of the posterior pdf for $h_0$ was 2.1 standard
deviations away from zero. For an initial detection of
gravitational radiation it is likely that the scientific community
will demand a significantly larger signal-to-noise ratio. However, the
performance of the MCMC routine is still very good for these relatively
low signal levels.

Although $~10^6$ Monte Carlo iterations were used in this study
(taking $~1$\,d on a 1\,GHz processor) adequate distributions can
be generated from $10^5$ iterations after the burn-in, so good
results can be achieved after just a few hours.  In fact the
marginalisations discussed above can be tackled more quickly
using simple summing methods as performed by \cite{Dupuis}, and
the result of a comparison of the two is shown in Fig.~\ref{fb}.
The great advantages of the MCMC method for us is its demonstrated
ability to deal with problems that have a large number of
parameters \cite{gilks96}, where other numerical integration
techniques (such as employed by \cite{Dupuis}) are not feasible.
The ultimate goal of our research is to expand this pulsar
parameter estimation work to include more parameters. The next
step in increasing the complexity of the pulsar signal is to
consider potential sources of known location, but with unknown
rotation frequency. In order to start this investigation we added
a new parameter, the uncertainty in the frequency of the source,
$\df$. In this example the exact value of the pulsar's
gravitational wave signal is uncertain to within $1/60$\,Hz. In
the study we present here there is a difference, $\df$, between
the gravitational wave signal frequency and the heterodyne
frequency. The addition of this new parameter did not
significantly increase the rate at which the code ran, but did (by
about $~20\%$) increase the length of the burn-in time. If one
wanted to increase this frequency range to $5$\,Hz then this
could be done by running the MCMC code
on 300 processors, with each run differing in center
frequency by $1/60$\,Hz. The Markov chain using the
\emph{correct} frequency would converge, while the other 299
chains would not. This will be a future research project for us.

\begin{figure}
\psfig{file=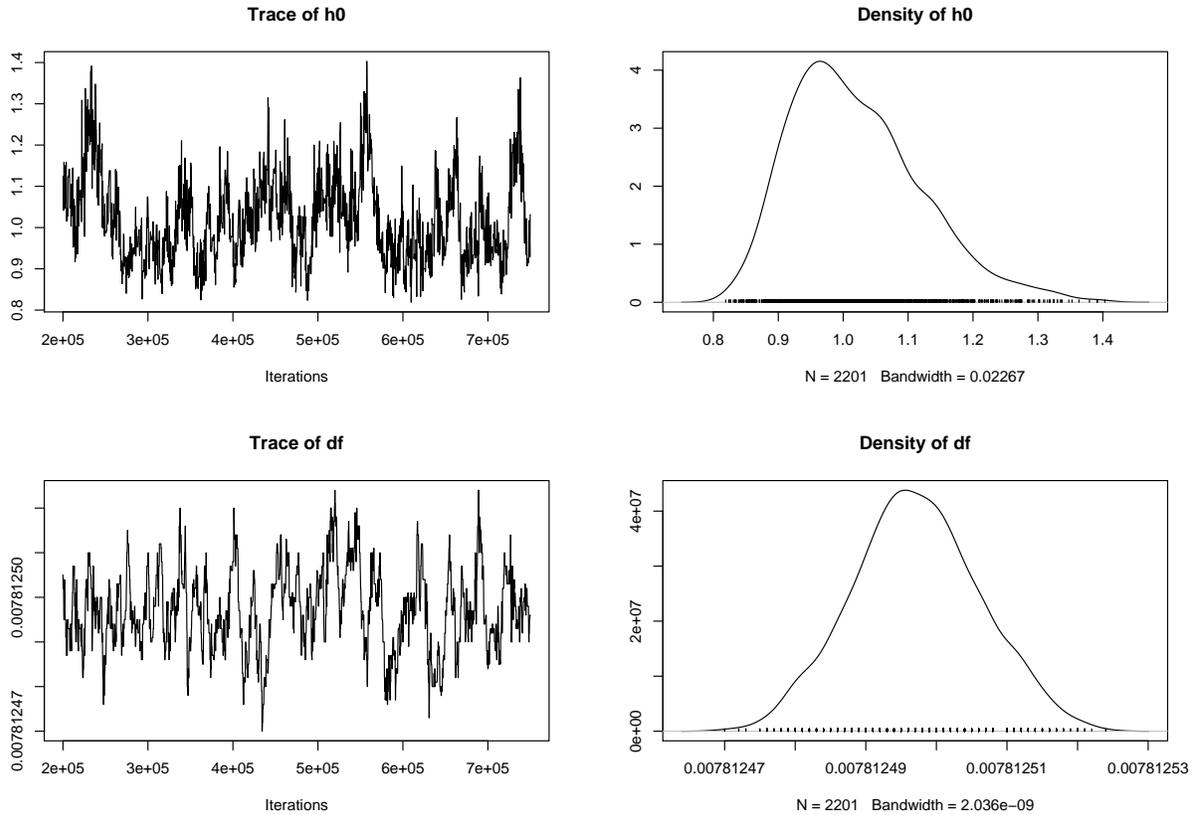,angle=-90,width=1.0 \linewidth,clip=}
\caption{\label{f3} Trace plot (left) and posterior pdfs (right)
for the pulsar parameters $h_0$ and $\df$. In this example from
the five parameter problem the true values for these critical
parameters were  $h_0=1.0$ and $\df=0.0078125$. }
\end{figure}

{
\psfrag{0}{0}\psfrag{1}{1}\psfrag{2}{2}\psfrag{3}{3}\psfrag{4}{4}
\psfrag{6}{6}\psfrag{8}{8}\psfrag{10}{10} \psfrag{0.0}{0.0}
\psfrag{0.5}{0.5} \psfrag{1.0}{1.0}
\psfrag{1.5}{1.5}
\begin{figure}
\includegraphics[width=5in]{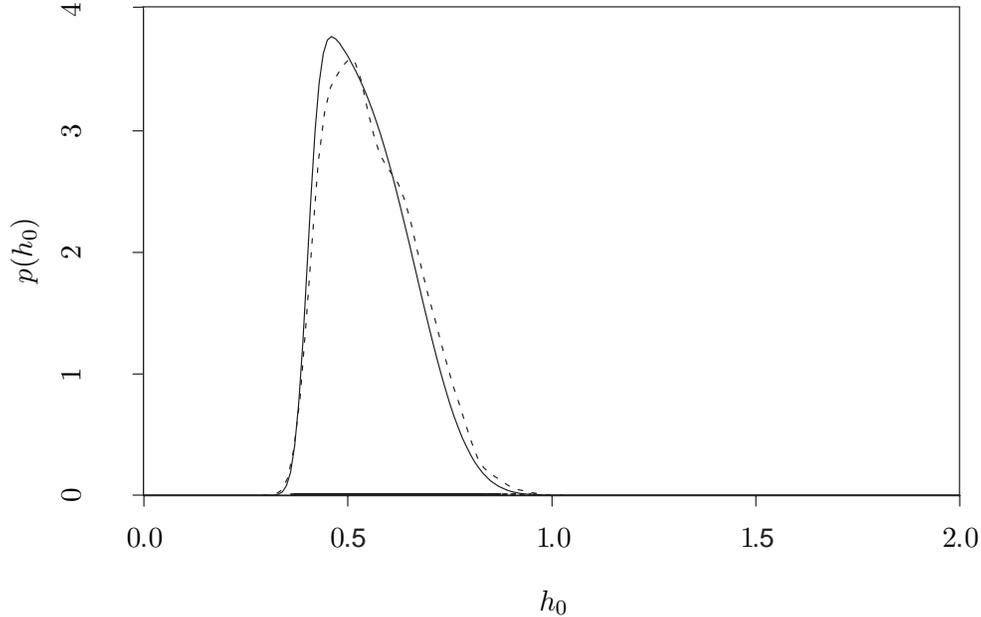}
\caption{\label{fb} The posterior mean based on the MCMC
sample for the gravitational wave amplitude parameter $h_0$
(dotted line), along with the that produced via the method
presented in \cite{Dupuis} (solid line) In this example the {\it true}
value was $h_0=0.5$, while the other {\it true} parameter values
were $\psi=0.4$, $\phi=1.0$, and $\iota=0.5$.}
\end{figure}
}
 In our MH code we used a uniform prior for the uncertainty in
the frequency, $\df$, over $\pm 0.016\,67$\,Hz. The injection
parameters used were $\psi=0.4$, $\phi=1.0$, $\df=0.007\,812\,5$,
and $\iota=0.5$ ($\cos\iota =0.88$). $h_0$ was again injected with
a number of values between $0.25$ and $10.0$. In Fig.~\ref{f3} we
show sample trace plots and posterior pdfs for $\Delta f$ and $h_0$
when the injected value of $h_0$ was $1.0$. For this example the
MCMC algorithm yielded mean values and 95\% posterior probability
intervals of $h_0=1.02$ ($0.86$ to $1.26$) and
$\df=0.007\,812\,497$ ($0.007\,812\,480$ to $0.007\,812\,515$).
The frequency pdf is quite narrow, which was responsible for the
increase in the burn-in time as the Markov chain must find this
narrow region of parameter space. In Fig.~\ref{f4} we display the
estimate for the gravitational wave amplitude ($h_0$) predicted by
the five parameter MH routine versus the actual $h_0$. In
Fig.~\ref{f5} we display the estimate for the
difference in frequency $\Delta f$
predicted by the five parameter MH routine versus the
injected $h_0$.

{ \psfrag{MCMC}{MCMC-derived $h_0$} \psfrag{Actual}{true $h_0$}
\psfrag{0}{0}\psfrag{1}{1}\psfrag{2}{2}\psfrag{3}{3}\psfrag{4}{4}
\psfrag{6}{6}\psfrag{8}{8}\psfrag{10}{10}
\begin{figure}
\includegraphics[width=5in]{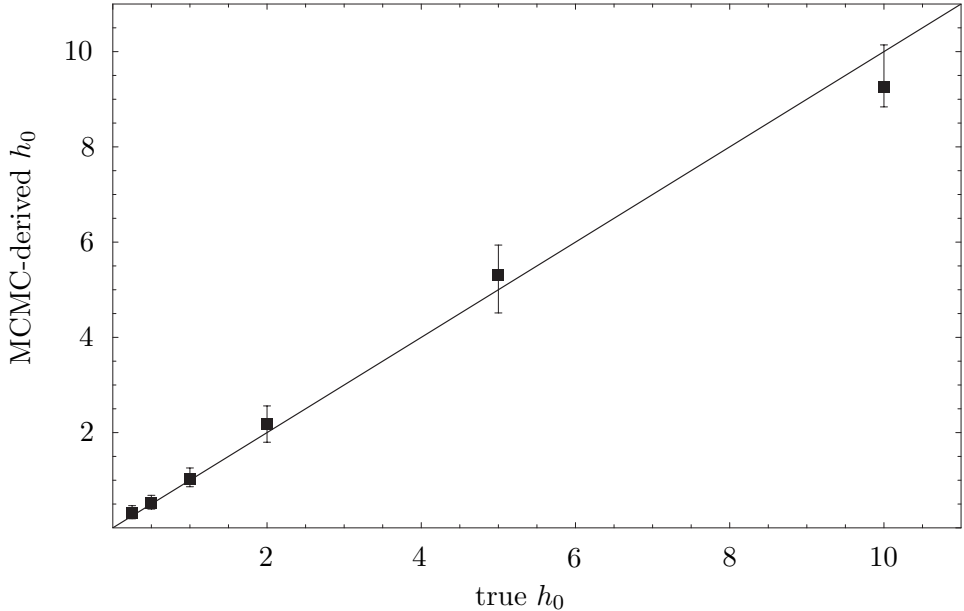}
\caption{\label{f4} The posterior mean based on the MCMC
sample for the gravitational wave amplitude parameter $h_0$ versus
the actual value of $h_0$ used in synthesizing the data. This
example is from the five parameter problem. The error bars
correspond to the lower and upper bound being specified by the
2.5\% and 97.5\% percentiles of the posterior pdf. The solid line
has a slope of $1$. }
\end{figure}
}

{ \psfrag{MCMC}{MCMC-derived $\df$ (Hz)} \psfrag{Actual}{true
$h_0$}
\begin{figure}
\includegraphics[width=5in]{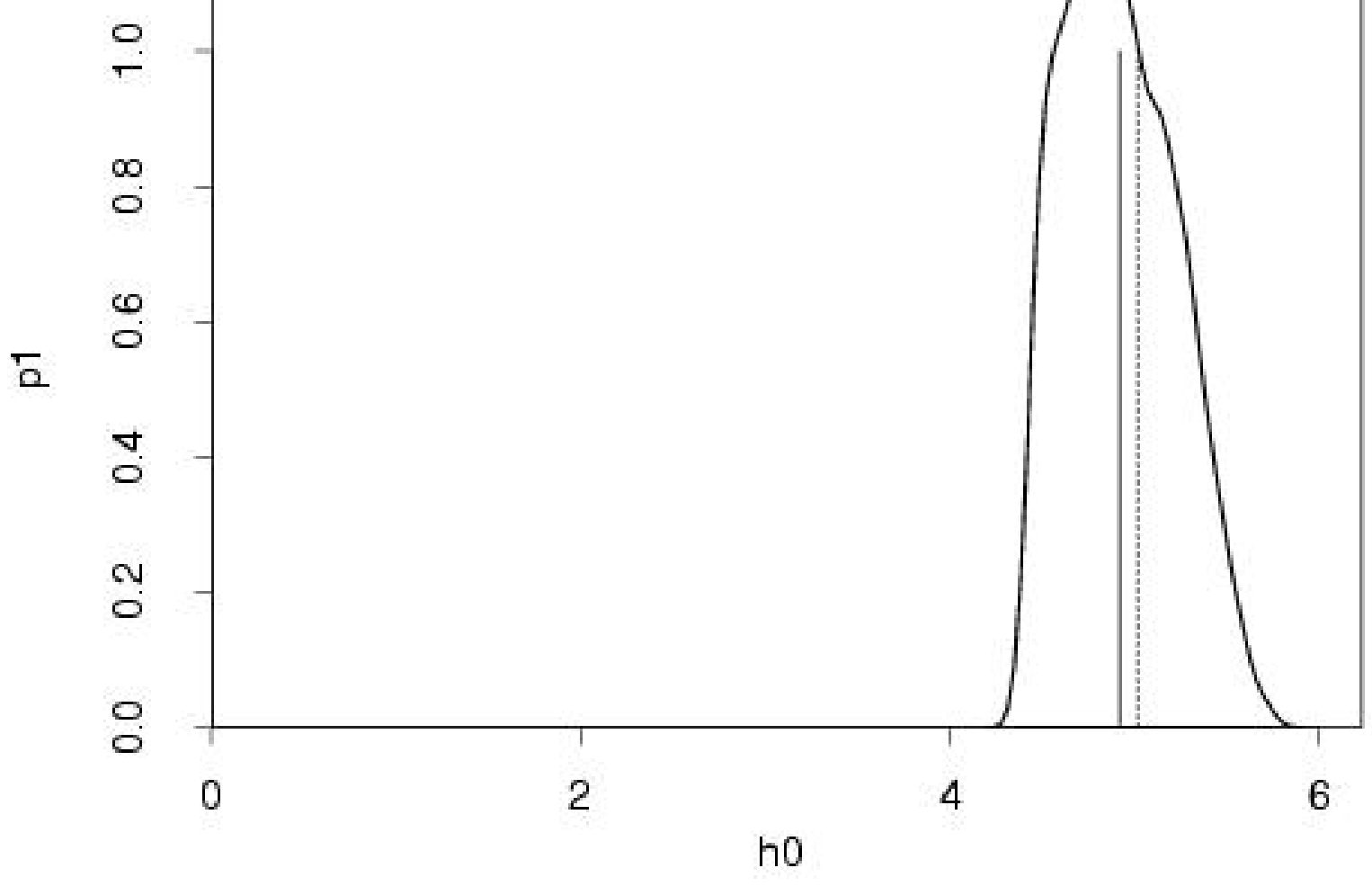}
\caption{\label{f5} The posterior mean based on the MCMC sample
for the uncertainty in the frequency, $\df$, versus the actual
value of $h_0$ used in synthesizing the data. This example is from
the five parameter problem. The error bars correspond to the lower
and upper bound being specified by the 2.5\% and 97.5\%
percentiles of the posterior pdf. The horizontal line corresponds
to the real value of $\df=0.007\,812\,500$. }
\end{figure}
}
\section{Discussion}
\label{disc}

Recent applications of MCMC techniques have provided a Bayesian
approach to estimating parameters in a number of physical
situations. These include cosmological parameter estimation from cosmic microwave
background data \cite{nlc1,knox,spergel}, estimating astrophysical
parameters for gravitational wave signals from coalescing compact
binary systems \cite{nlc2,nlc3}, and parameter estimation
of a chaotic system in the presence of noise \cite{meyer1,meyer2}.
An all sky survey for
periodic gravitational waves from neutron stars must explore
a very large parameter, and this has partially been addressed in \cite{jaranowski}.
Generically, the signal from a neutron star in a binary system
will be characterized by at least 13 parameters.
It is our
hope that MCMC techniques will prove fruitful in dealing with
these complex signals.

In this paper we have demonstrated that the success of MH routine
for the five parameter problem: $h_0$, $\psi$, $\phi_0$, $\iota$
and $\df$. Our longer term plans are to account for other
parameters, such as spindown rate, pulsar wobble, and possibly
location of the signal in the sky. This research is currently in
progress.

\begin{acknowledgments}
This work was supported by National Science Foundation grants
PHY-0071327 and PHY-0244357, the Royal Society of New Zealand
Marsden fund award UOA204, the Natural Sciences and Engineering
Research Council of Canada, Universities UK, and the University
of Glasgow.
\end{acknowledgments}

\end{document}